\documentclass[aps,twocolumn,showpacs]{revtex4}
\usepackage{amssymb}
\usepackage{amsmath}
\usepackage{epsfig}

\begin{document}

\title{Spin Hall effect at interfaces between HgTe/CdTe quantum wells and metals}

\author{M. Guigou$^{1,2}$, P. Recher$^{2}$, J. Cayssol$^{1,3,4}$, and B. Trauzettel$^{2}$}

\affiliation{$^{1}$ Condensed Matter Theory Group, LOMA (UMR-5798), CNRS and University Bordeaux 1, F-33045 Talence, France\\
$^2$ Institut f\"ur Theoretische Physik und Astrophysik, University of W\"urzburg, 97074 W\"urzburg, Germany\\
$^3$ Department of Physics, University of California, Berkeley, California
94720, USA\\
$^4$ Max-Planck-Institut f\"ur Physik Komplexer Systeme, D-01187 Dresden, Germany}

\begin{abstract}
We study the spin-dependent transmission through interfaces between a HgTe/CdTe quantum well (QW) and a metal for both the normal metal and the superconducting cases. Interestingly, we discover a type of spin Hall effect at these interfaces that happens to exist even in the absence of structure and bulk inversion asymmetry within each subsystem (i.e., the QW and the metal). Thus, this is a pure boundary spin Hall effect which can be directly related to the existence of exponentially localized edge states at the interface. We demonstrate how this effect can be measured and functionalized for an all-electric spin injection into normal metal leads.
\end{abstract}


\maketitle

\section{Introduction}

The Spin Hall effect (SHE) is a rich physical phenomenon which, since its prediction \cite{Dyakonov1971}, has led to many interesting discoveries in the field of spintronics. Realized in nonmagnetic systems, the SHE allows for an all-electrical manipulation of spin. The underlying interaction of the SHE is spin-orbit coupling with the observable consequence that a transverse spin current is generated if an electrical charge current is driven in the longitudinal direction. Notably, this can happen due to impurity scattering \cite{Dyakonov1971,Hirsch1999}, the {\it extrinsic} SHE, or due to band structure effects \cite{Murakami2003,Sinova2004}, the called {\it intrinsic} SHE. The latter case is usually directly related to either structure inversion or bulk inversion asymmetry of the underlying nanostructure, resulting in Rashba or Dresselhaus spin-orbit coupling, respectively.

Remarkably, by now, both cases have been experimentally observed: the {\it extrinsic} SHE in semiconductor heterostructures by optical \cite{Kato2004} and electronic \cite{Garlid2010} measurement techniques and the {\it intrinsic} SHE in two-dimensional semiconductors \cite{Wunderlich2005} as well as in HgTe/CdTe heterostructures by combining the SHE with the so-called quantum spin Hall effect (QSHE) in a single device \cite{Brune2010,Brune2011}. The QSHE is yet another type of spin Hall effect that exists at the boundary of a two-dimensional topological insulator realized in HgTe/CdTe quantum wells (QWs) \cite{Konig2007}. It refers to the existence of protected metallic edge states propagating in opposite directions and forming a single set of Kramers partners at each edge \cite{Kane2005,Bernevig2006,Fu2007}. The topological phase transition behind this effect is related to an inversion of bands with opposite parities due to the strong spin-orbit interaction and it is well described by a massive Dirac equation \cite{Bernevig2006}. The critical point is then reached when the mass of the Dirac fermions goes to zero. Directly at that point, the system behaves like a single valley Dirac fermion; this has been experimentally confirmed \cite{Buttner2010}.

Recently, Yokoyama and co-workers predicted a giant spin rotation at a junction between a normal metal and QSHE system \cite{Yokoyama2009}; this is one of the main motivations of our work. Here, we go substantially beyond this prediction in two different ways. First and most importantly, we discover a different type of interface SHE at junctions between HgTe/CdTe QWs and metals (for both the normal metal and the superconducting cases). Second, we functionalize our findings to propose a device for all-electric spin injection into normal metal leads in the absence of ferromagnetic contacts. All our predictions apply to nanostructures in the ballistic transport regime.

We show below that this type of SHE is intimately related to the coexistence of propagating and evanescent modes at the interface between a QSHE system and a metal. Mathematically, this comes from the fact that the underlying low-energy Hamiltonian contains terms that are linear {\it and} quadratic in the electron wave vector. Physically, this interplay can happen near a band-inversion crossing that in the case of HgTe/CdTe QWs drives the topological phase transition \cite{Bernevig2006}. Interestingly, this effect exists even in the total absence of structure and bulk inversion asymmetry within each subsystem. To the best of our knowledge, this is the first prediction of a SHE in a composite system that does not break these symmetries in either subsystem.

\begin{figure}[h]
\begin{center}
\includegraphics[width=8cm]{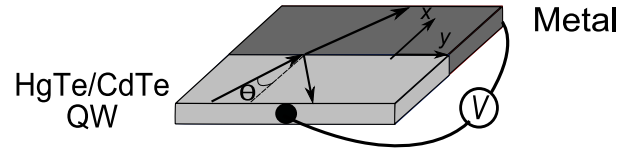}
\end{center}
\vspace*{-0.1cm}
\caption{Junction between a HgTe/CdTe QW (light gray) and a metal (dark gray). Due to the applied voltage $V$, electrons are injected from a contact with an angle of incidence $\theta$ with respect to the interface. At the interface they are either reflected or transmitted.}
\label{system}
\end{figure}

The paper is organized as follows. In Sec. II, we present the model Hamiltonian for a HgTe/CdTe quantum well 
and describe the formalism needed to compute the scattering probabilities of a HgTe QW/metal junction.
In Sec. III, we investigate the spin- and angular-dependence of transmission as well as the Andreev reflection probabilities 
and analyze their asymmetric behavior as an interfacial spin Hall effect. We propose to measure this effect and functionalize it by an experimental realization 
described in Sec. IV. We conclude in Sec. V. Details of calculations are given in the Appendix.

\section{Model}
This section is devoted to the theoretical framework of the junction between a HgTe/CdTe quantum well and a metal. 
We present the model Hamiltonian for the HgTe/CdTe quantum well and describe the scattering matrix method which allows us to calculate to the scattering probabilities through the interface.
\subsection{Hamiltonian}
The band structure of a HgTe/CdTe QW is derived from the eight-band Kane model \cite{Novik2005}. Near the topological phase transition, it can be described by an effective four-band Hamiltonian with two subbands commonly called $E1$ and $H1$ that have opposite parities. The $E1$ and the $H1$ subbands are both doubly degenerate due to time-reversal symmetry (TRS). We refer to these degenerate Kramers partners as spin $\uparrow,\downarrow$.  Near the $\Gamma$ point, the effective Hamiltonian can be written as \cite{Bernevig2006}
\begin{equation}
H(\mathbf{k})=%
\begin{pmatrix}
h(\mathbf{k}) & 0 \\
0 & h^{\ast }(-\mathbf{k})%
\end{pmatrix}%
,  \label{BHZ}
\end{equation}%
where $h(\mathbf{k})=\epsilon(\mathbf{k})I_{2\times2}+d_a(\mathbf{k})\sigma^a$ represents the spin-$\uparrow$ block in the ($\left\vert E1\uparrow\right\rangle ,\left\vert
H1\uparrow\right\rangle$) space, $h^{*}(-\mathbf{k})$ the spin-$\downarrow$ block in the ($\left\vert E1\downarrow\right\rangle ,\left\vert
H1\downarrow\right\rangle$) space, and $\sigma^a$ denote the Pauli matrices. In Eq.~($\ref{BHZ}$), $\epsilon(\mathbf{k})=C-Dk^2$, $d_a(\mathbf{k})=[Ak_x,-Ak_y,M(\mathbf{k})]$, and $M(\mathbf{k})=M-Bk^2$, where $\mathbf{k}=(k_x,k_y)$ is the in-plane momentum, $k^2=k_x^2+k_y^2$ and $M$ represents the Dirac mass. The parameters $A$, $B$, $C$, $D$, and $M$ depend on the geometry of the HgTe/CdTe QW. The sign of $M$ relative to $B$ distinguishes the trivial phase ($M>0$) from the non-trivial phase ($M<0$, $B$ being negative). The eigenenergies of $h(\mathbf{k})$ are given by $E_{\pm}=\epsilon(\mathbf{k})\pm d(\mathbf{k})$ where $d(\mathbf{k})=\sqrt{A^2k^2+M(\mathbf{k})^2}$ and $\pm$ refers to the conduction (valence) band.

\subsection{Scattering method}

We consider a junction between a HgTe/CdTe QW and a normal metal or an $s$-wave superconductor as depicted in Fig.~$\ref{system}$. The interface is assumed to be perfect and located at $x=0$. The model of this junction relies on a step-like variation of the bands, modeled by $C=C_L$ for $x<0$ and $C=C_R$ for $x>0$. We make the reasonable assumption to model the normal metal as a highly doped HgTe/CdTe QW (similar to the treatment of the corresponding problem in graphene \cite{Tworzydlo2006}) and the proximity-induced superconductivity with a step-like varying pair potential \cite{Beenakker2006}. We assume that the only boundary of the problem is the QW/metal interface. Therefore, we analyze bulk state transport in the remainder of this paper.

The scattering states in the bulk of the HgTe/CdTe QW can be written for each block of the Hamiltonian (\ref{BHZ}) separately. 
They are plane wave (two-component) spinor wave functions that depend on all the parameters of the Hamiltonian, in particular on $k_x$ and $k_y$. 
Since we assume translation invariance in the $y$ direction, the transverse wave vector $k_y$ is conserved. 
Due to the block structure of the Hamiltonian, we can analyze the spin-$\uparrow$ scattering problem separately from the spin-$\downarrow$ case.
 In the QW region ($x<0$), the spin-up quasiparticles are described by two-components spinors \cite{Yokoyama2009,Novik2009,LBZhang2009}
\begin{equation}
\Psi(x)=\left(
\begin{array}{c}
\pm d(\mathbf{k})+M(\mathbf{k}) \\ 
A(k_x-ik_y)%
\end{array}%
\right) e^{ik_xx},
\label{spinor}
\end{equation}
where the $\pm$ sign labels the conduction (valence) subband. The dispersion relation $E(k)$ yields two possible wave vectors, namely
\begin{eqnarray}
k_{1,2} &=&\Bigg(\frac{\gamma\pm \sqrt{\gamma^2-\eta} }{2(B^{2}-D^{2})}\Bigg)^{1/2}
\label{expressk}
\end{eqnarray}
 with $\gamma =-A^{2}+2MB-2D(C_{L}-E)$ and $\eta=4(B^2-D^2)[M^2-(C_L-E)^2]$. 
We make the further assumptions that $B^{2}>D^{2}$ and $M<\frac{A^2-2D(E-C_L)}{2B}$, which are motivated by typical parameters for HgTe/CdTe QWs. 
Under this choice of parameters, we find that $k_1$ is real and $k_2$ is imaginary. 
This results in the coexistence of propagating and evanescent modes on both sides of the boundary, which gives rise to the interface SHE further discussed below. 
On the left-hand side, propagating electrons have a real wave vector $k_x= \pm k_1 \cos \theta$, where $\theta$ is the angle of incidence and 
$\pm$ labels the incident (reflected) mode.
Meanwhile, evanescent electrons are described by a complex wave vector $k_x= -i(k_y^2-k_2^2)^{-1/2}$.


At the metal side of the interface ($x>0$), the outgoing scattering states have a similar form as the ones for $x<0$ but the definitions 
 of the wave vectors $k_{1,2}$ contain the parameter $C_R$ instead of $C_L$ (see Appendix 1 for more details). 
To obtain the scattering amplitudes for transmission through the junction, we match the wave functions and their derivatives at $x=0$. 
Once this calculation has been done for the spin-$\uparrow$ block the corresponding spin-$\downarrow$ problem follows by TRS.


\section{Numerical results and discussion}

In this section, we investigate the scattering probabilities, which depend on the mass $M$ and angle of incidence $\theta$. 
We first focus on the junction between a HgTe QW and a normal metal for two different doping configurations. 
Afterward, we consider the contact of a HgTe QW and an $s$-wave superconductor involving Andreev reflection processes. 
In both cases, the angular asymmetric behavior of the scattering probabilities is connected to the presence of evanescent modes. 
The coexistence of evanescent and propagating waves is necessary for the appearance of an interfacial spin Hall effect. 

\subsection{Normal metal case}

\begin{figure}[h!]
\includegraphics[width=3.8cm]{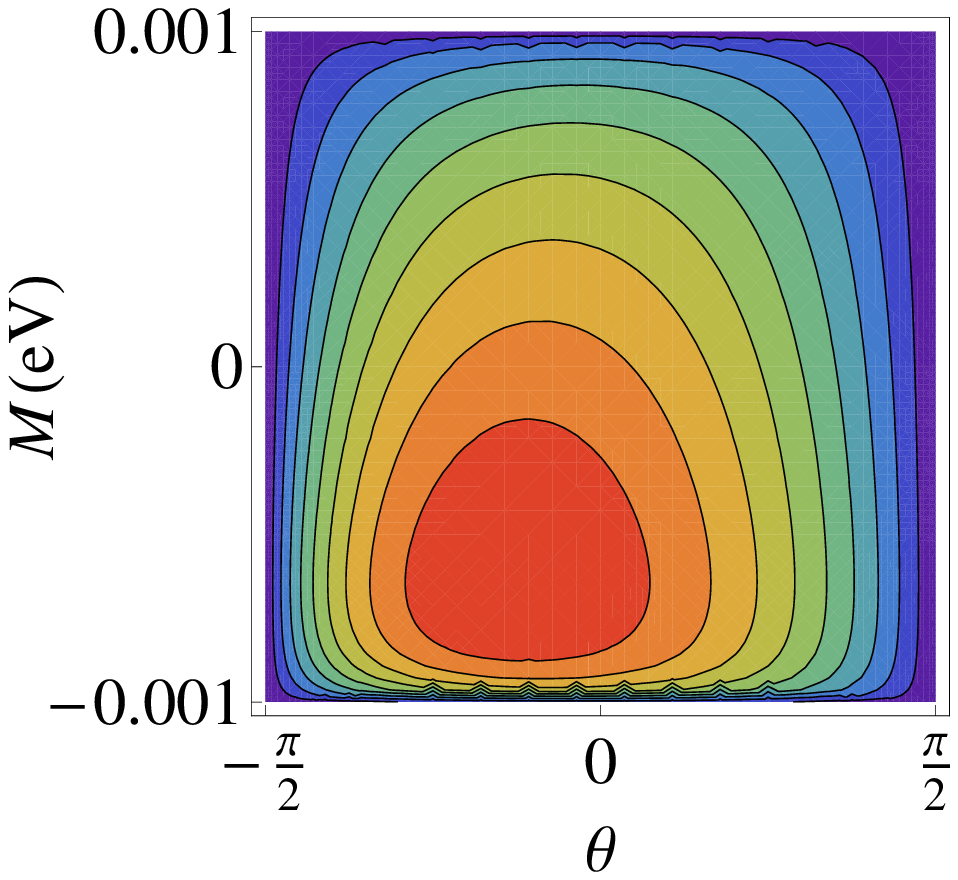}
\includegraphics[width=.5cm]{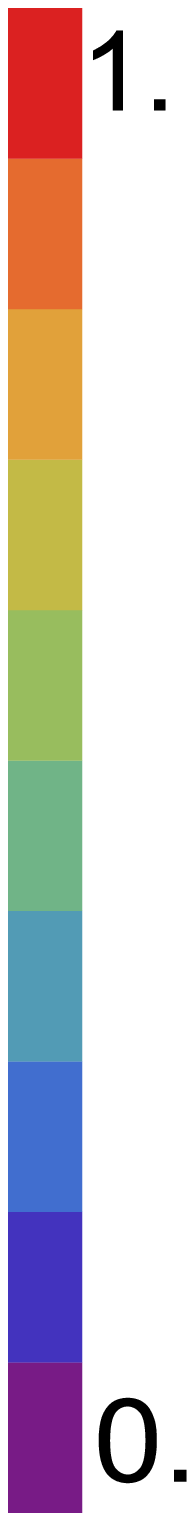}
\includegraphics[width=3.8cm]{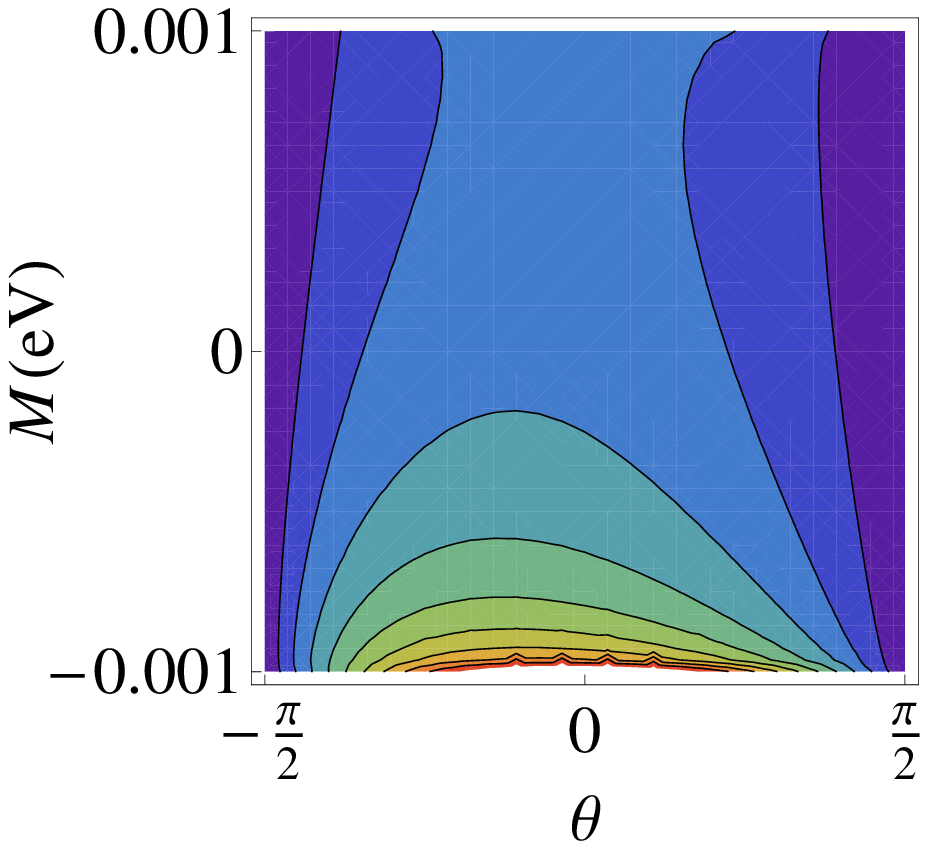}
\includegraphics[width=.5cm]{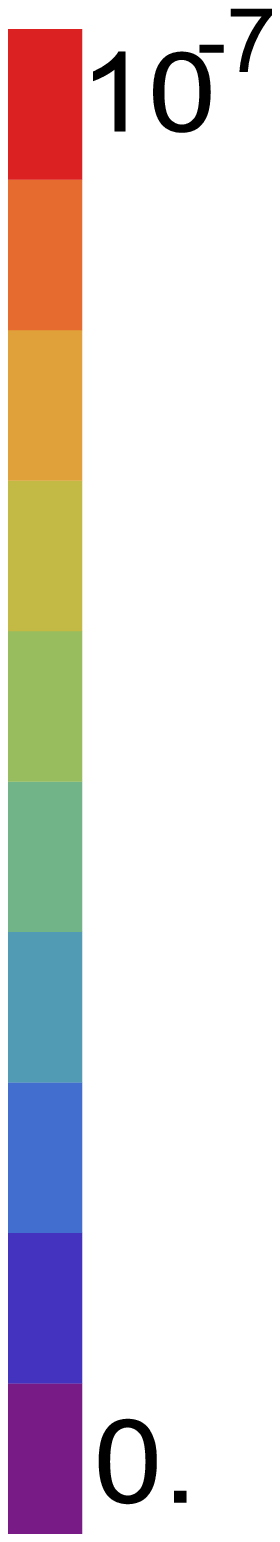}\hspace*{-0.3cm}
\caption{(Color online) Angle dependence of the transmission probability (left panel) and the amplitude squared of the evanescent mode 
on the metal side (right panel) for the spin-$\uparrow$ block for a junction between a low-$n$-doped HgTe/CdTe QW  and an $n'$-doped metal at the Fermi
 level ($E=0$) as functions of the mass parameter $M$, where $-C_L=1$ $\mathrm{meV}$ and $-C_R=500$ $\mathrm{meV}$. The other parameters are those found 
in typical HgTe QW experiments, i.e., $A=4$ $\mathrm{eV}$ $\mathrm{\mathring{A}}$, $B=-70$ $\mathrm{eV} $ $\mathrm{\mathring{A}}^{2}$ and $D=-50$ $\mathrm{eV}$ $\mathrm{\mathring{A}}^{2}$. A clear asymmetry is present for negative values of $M$.}
\label{nncase}
\end{figure}

We first consider the scattering amplitudes at the Fermi energy $E=0$  through a HgTe/CdTe QW-normal metal junction when both subsystems are $n$-doped. 
The transmission probability and the amplitude squared of the evanescent mode on the right-hand side of the junction are shown in Fig.~$\ref{nncase}$ 
as functions of the mass parameter $M$ and the angle of incidence $\theta$.

For positive values of the gap $M$, the transmission probability exhibits a symmetric behavior with respect to $\theta$ and has a rather weak amplitude.
 On the contrary, for a large negative mass $M$, this amplitude increases and tends to peak at negative $\theta$.
 Comparing the transmission with the amplitude squared of the evanescent mode (see the right panel of Fig.~$\ref{nncase}$),
 a correlation between the two is visible. Hence, the magnitude of the amplitude of the evanescent mode is connected to the angle and spin dependence
 of the transmission. (Note that the spin-$\downarrow$ case just follows from the substitution $\theta \rightarrow -\theta$ in the results shown in Fig.~$\ref{nncase}$.
) This gives rise to a spin current at the interface in the transverse direction \cite{foot_rothe}. Only the coexistence of propagating and evanescent 
modes at the interface gives rise to this effect. In the parameter regime where evanescent modes are absent, one can easily show that the transmission of the two spin blocks has to be symmetric with respect to $\theta$. We remark in passing that the $p-n'$ case yields similar results to the $n-n'$ case discussed here.
\begin{figure}[h!]
\includegraphics[width=3.8cm]{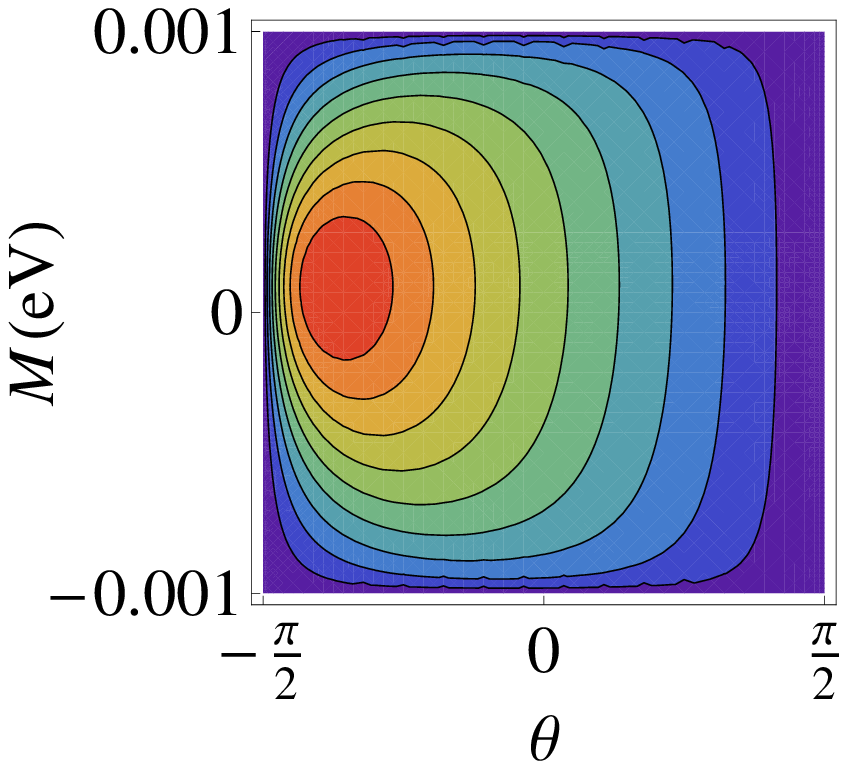}
\includegraphics[width=.5cm]{fig3.eps}
\includegraphics[width=3.8cm]{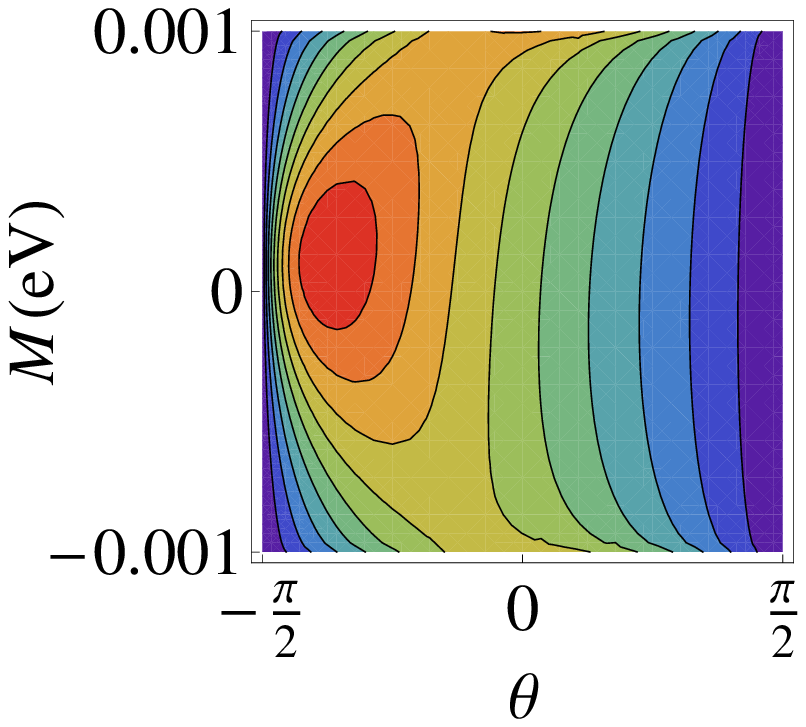}
\includegraphics[width=.5cm]{fig5.eps}\hspace*{-0.3cm}
\caption{(Color online) Angle dependence of the transmission probability (left panel) and the amplitude squared of the evanescent mode on 
the metal side (right panel) for the spin-$\uparrow$ block for a junction between a low-$n$-doped HgTe/CdTe QW  and a $p$-doped metal at the Fermi level 
($E=0$) as functions of the mass parameter $M$, where $-C_L=1$ $\mathrm{meV}$ and $C_R=500$ $\mathrm{meV}$. The values of the parameters $A$, $B$, and $D$ are the same as in Fig.~$\ref{nncase}$. Now, a clear asymmetry is present for values of $M$ close to the critical point.}
\label{npcase}
\end{figure}

Interestingly, when the metal side is $p$-doped (and not $n'$-doped as discussed before in Fig.~$\ref{nncase}$), a similar asymmetry in the spin- and angle- dependent transmission holds, but the maximum then appears near the critical point $M\rightarrow 0$ (see Fig.~$\ref{npcase}$).

\subsection{Superconducting case}

We now turn to the analysis of a junction between a HgTe QW and a superconductor (SC). 
Then, the Hamiltonian of Eq.~($\ref{BHZ}$) must be completed by particle-hole symmetry and the pairing potential matrix $\Delta(\mathbf{k})$. 
$[$We assume $s$-wave singlet pairing; see Eq.~($\ref{BDG}$) in the Appendix.$]$ In our model, the HgTe/CdTe QW on the left-hand side of the interface ($x<0$)
 with a chemical potential $C_L$ has no pairing potential, 
while the Hamiltonian of the SC side ($x>0$), which may have a different electronic filling $C_R$, has to be supplemented by a finite (proximity-induced)
 order parameter $\Delta(\mathbf{r})=\Delta_0e^{i\phi}$, where $\phi$ is the superconducting phase. 
The scattering amplitudes of such an interface can be calculated along the lines of Ref.~\cite{Guigou2010} (details are given in Appendix 2). Here, we restrict ourselves to subgap transport $[$based on Andreev reflection (AR)$]$ where the quasiparticle excitation energy $\epsilon$ is smaller than the superconducting gap $\Delta_0$.
\begin{figure}[h!]
\includegraphics[width=3.8cm]{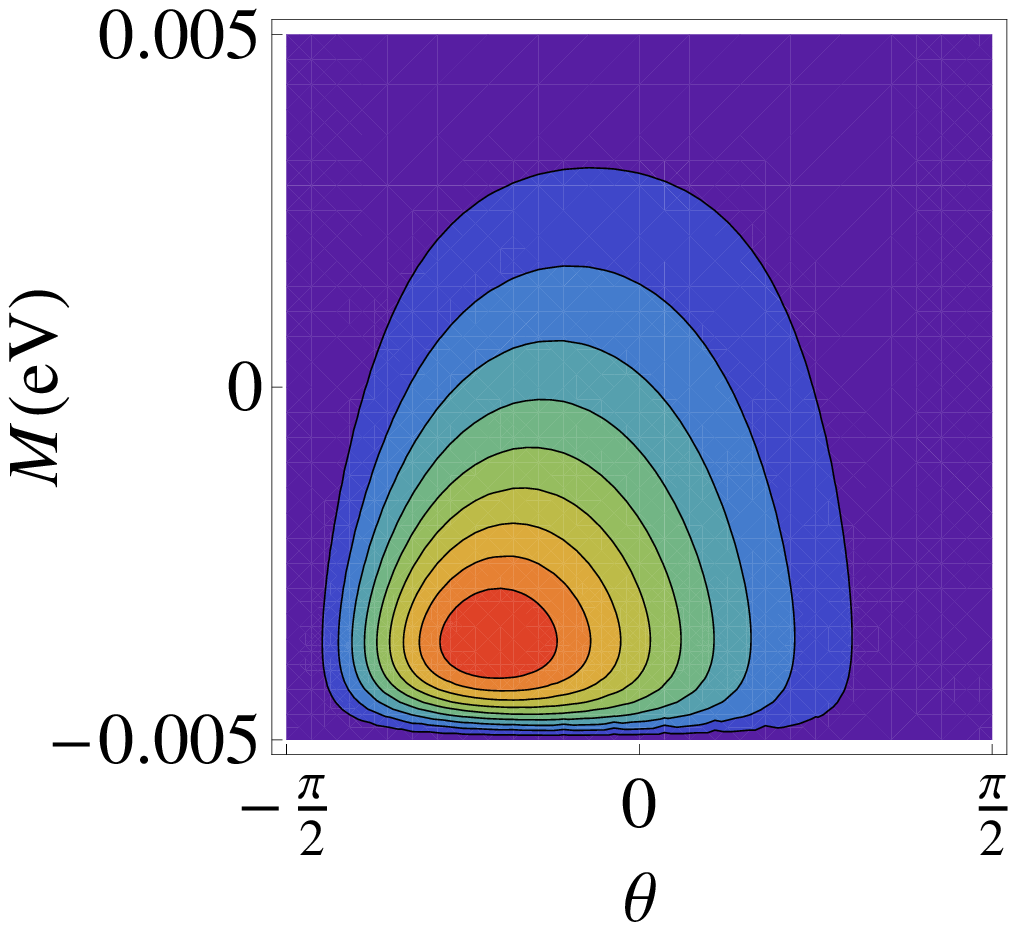}
\includegraphics[width=0.5cm]{fig3.eps}
\includegraphics[width=3.8cm]{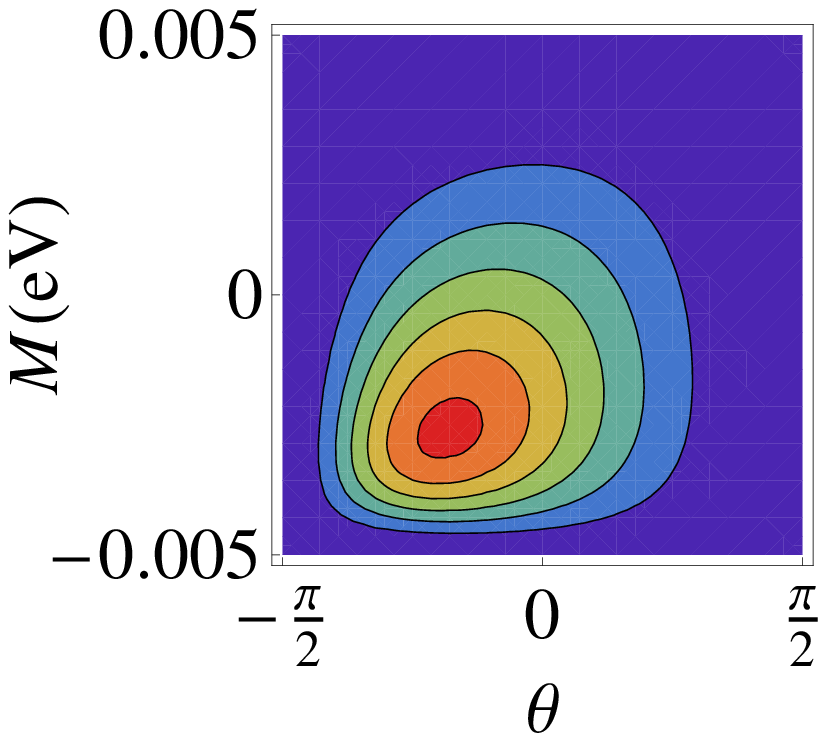}
\includegraphics[width=0.5cm]{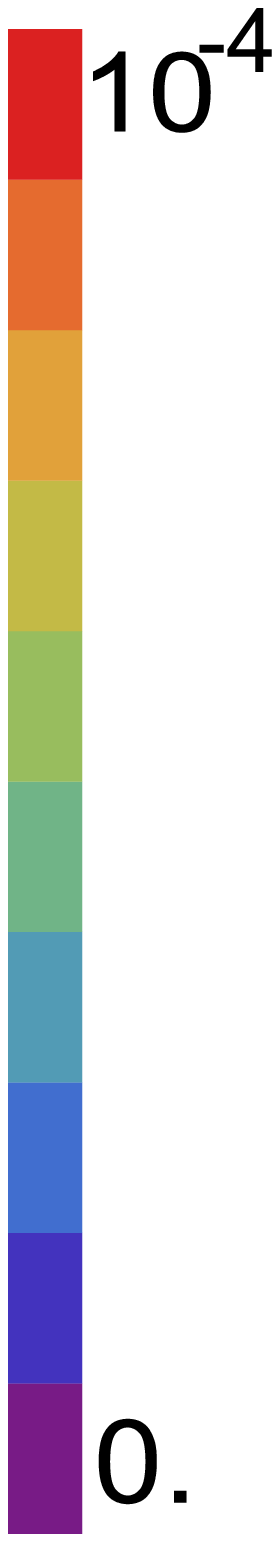}\hspace*{-0.3cm}
\caption{(Color online) Angle dependence of the AR probability (left panel) and the amplitude squared of the evanescent (hole-like) mode on 
the HgTe/CdTe QW side (right panel) for an incident spin-$\uparrow$ electron at a junction between a low-$n$-doped HgTe/CdTe QW and a SC at the Fermi 
level ($\epsilon=0$) as functions of the mass parameter $M$, where $-C_L=5$ $\mathrm{meV}$, $-C_R=1$ $\mathrm{eV}$ and $\Delta_0=1$ $\mathrm{meV}$. 
The values of the parameters $A$, $B$, and $D$ are the same as in Fig. $\ref{nncase}$. Similarly to Fig. $\ref{nncase}$, a clear asymmetry is present for negative values of $M$.}
\label{figure2}
\end{figure}

In Fig.~$\ref{figure2}$, we show the behavior of the AR probability and the weight of evanescent modes on the left-hand side for an incident 
spin-$\uparrow$ electron as a function of $M$ and the angle of incidence $\theta$ at the Fermi energy. Similarly to the $n-n'$ junction, (see Fig.~$\ref{nncase}$),
 the AR exhibits an asymmetric behavior with respect to $\theta$, with a maximum for large negative $M$. Again, an evident correlation between the presence of evanescent modes and the asymmetry is observed; compare the left and the right panels of Fig.~$\ref{figure2}$ with each other. The case of an incident spin-$\downarrow$ electron just follows by replacing $\theta \rightarrow -\theta$ in Fig.~$\ref{figure2}$.

We remark here that raising the electronic filling of the HgTe/CdTe QW at the left-hand side of the interface alters substantially the asymmetric
 behavior of the scattering amplitudes in both types of junction (normal metal and SC). As expected from a junction between two metals, asymmetry in
 such a scattering problem is absent. Strong asymmetry is also absent when the linear
terms are dominant, as in graphene \cite{Tworzydlo2006,Beenakker2006} or low-doped HgTe/CdTe wells \cite{LBZhang2009}. That is the reason that the
effect has been missed before.

Hence, the simultaneous existence of propagating and evanescent modes leads to the appearance of a type of spin Hall effect manifested as a
 local spin current density flowing along the interface. A similar SHE was recently predicted at potential steps in the Kane-Mele model in graphene
 \cite{Yamakage2010} and on the surface of a three-dimensional topological insulator \cite{Gao2011}.
\begin{figure}
\begin{center}
\epsfxsize4.3cm \epsffile{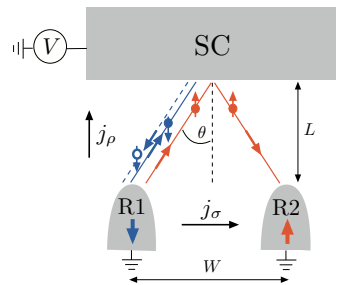}
\end{center}
\par
\vspace*{-0.2cm}
\caption{(Color online) Schematic of the AR spin splitter: Electrons are injected from an unpolarized normal metal (point-like) reservoir R1 (and similarly for R2) due to a bias voltage $V$. As discussed before, we can adjust parameters such that electrons with spin-$\downarrow$ (blue full line) arrive at the junction with an angle of incidence favorable for AR. Hence, they are transmitted as Cooper pairs into the SC. This process produces reflected holes with spin-$\downarrow$ collected into the first reservoir R1 (blue dashed line). At the same time, this angle of incidence is less favorable for particle-hole conversion of electrons with spin-$\uparrow$ (red full line). Thus, these electrons are mostly reflected at the interface to a second reservoir (R2), resulting in a spin imbalance between R1 and R2. Hence, there will be a finite charge current $j_\rho$ from the reservoirs R1 and R2 to the SC and there will be a finite spin current $j_{\sigma}$ from R1 to R2. This spin current yields a spin accumulation in the two reservoirs in the direction of the bold arrows.}
\label{setup}
\end{figure}

However, the junctions based on HgTe/CdTe QWs discussed in our work are, to the best of our knowledge, the first composite system where a SHE can be observed in the absence of structure and bulk inversion asymmetry within each subsystem.

\section[test]{Experimental realization}

The observation of such an effect along the interface is complicated although the use of spin-resolved scanning tunneling microscopy techniques or
optical methods might work \cite{Werake2011}. Here, we propose an alternative way, depicted in Fig. $\ref{setup}$, in order to take advantage
 of the spin- and angle- dependence of the AR probability. When a voltage $V$ is applied across the QW/SC contact, electrons are injected through the junction from normal metal (point-like) reservoirs labeled R1, R2 in the figure. Then, as discussed in detail above, the presence of evanescent modes showing up at the interface acts on the incoming wave as a spin splitter whose efficiency depends on various parameters. A detailed description of the working principle of the envisioned device is given in the caption of Fig.~\ref{setup}.

To quantify the efficiency of the spin splitter, we calculate the spin conductance, which is a direct measure of the imbalance between
 spin-$\uparrow$ and spin-$\downarrow$ carriers in reservoirs R1 and R2, induced by the voltage $V$. It can be written as
\begin{eqnarray}
G_S (\theta,V) \equiv \frac{2e^2}{h}\sum_{\sigma=\pm}\sigma|r_{ee,\mathbf{\sigma}}|^2,
\label{DSC}
\end{eqnarray}
where $|r_{ee,\mathbf{\sigma}}|^2$ is the probability for an electron with spin $\sigma$ and injected from R1 at an angle $\theta=\arctan(W/2L)$ to be
 reflected at the interface. $W$ and $L$ are respectively the width of the system and the distance between the reservoirs and the interface. The factor $2$ accounts for the two reflection processes, from R1 to R2 and vice versa.
\begin{figure}
\vspace*{0.1cm}
\begin{center}
\epsfxsize6.cm \epsffile{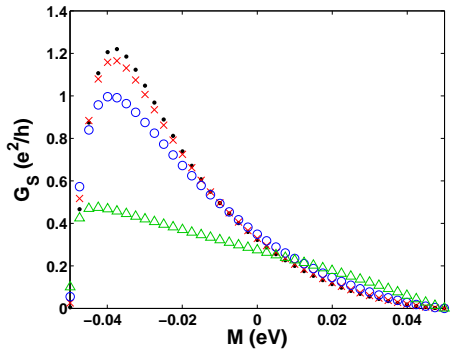}
\end{center}
\par
\vspace*{-0.2cm}
\caption{(Color online) Bias dependence of the differential spin conductance for a QW/SC contact with $-C_L=50$ $\mathrm{meV}$ and $-C_R=1$ $\mathrm{eV}$.
 $G_{S}$ is plotted for different values of the injection energy: $eV=0.$ (black dots); $0.3$ $\Delta_0$ (red crosses); $0.6$ $\Delta_0$ (blue circles), and $0.9$ $\Delta_0$ (green triangles).
 The other parameters are $\Delta_0=1$ $\mathrm{meV}$, $A=4$ $\mathrm{eV}$ $\mathrm{\mathring{A}}$, $B=-70$ $\mathrm{eV}$ $\mathrm{\mathring{A}}^{2}$, and $D=-50$ $\mathrm{eV}$ $\mathrm{\mathring{A}}^{2}$. The size of the system is taken as $W=500$ $\mathrm{nm}$ and $L=500$ $\mathrm{nm}$. It is clearly visible that the spin splitter works best at a small bias voltage.}
\label{pola_M}
\end{figure}

We mention here that the contribution to Eq.~($\ref{DSC}$) from topologically protected edge states (if the QW has a finite width $W$) is negligible. Hence, our result also applies to finite but wide systems where the overlap between edge states at opposite edges is weak. In Fig.~$\ref{pola_M}$, we additionally analyze the energy and band gap dependence of the performance of the AR spin splitter. It is clearly visible that the device works best at small bias voltages and that the ideal value of the band gap $M$ changes slightly to more negative values as the bias is increased.
\section[conclusion]{Conclusion}

To summarize, we have predicted spin-dependent transport properties at interfaces between HgTe/CdTe QWs and normal metals as well as superconductors, 
resulting in an interface spin Hall effect. It has been identified that this effect is clearly connected to the coexistence of propagating and evanescent modes at the junction. We have proposed a setup to functionalize it as an Andreev reflection spin splitter.

\section*{ACKNOWLEDGMENTS}

We would like to thank the DFG-JST research unit ``Topological Electronics'' (M.G. and B.T.), the Emmy-Noether program (P.R.), 
and the PROCOPE program for financial support. J.C. acknowledges support from EU/FP7 under the contract TEMSSOC and from ANR through Project No. 2010-BLANC-041902 (ISOTOP). 
Futhermore, we acknowledge interesting discussions with E. M. Hankiewicz and D. G. Rothe.

\section*{APPENDIX}

In this Appendix, we write explicitly the wave functions involved in a junction between a HgTe/CdTe QW and a normal metal or an $s$-wave superconductor. 
Then, we define the scattering probabilities plotted in Figs. 2-4. 

We solve the scattering problem for an incoming spin-up electron in the conduction band. All the states are plane waves and are written as $\Psi(x)e^{ik_yy}$, where $\Psi(x)$ is a two-component spinor wave function. 

\subsection*{1. HgTe/CdTe QW/metal interface}
We first consider the interface between a HgTe/CdTe QW and a metal modeled as a highly doped HgTe/CdTe QW.

On the left-hand side of the interface, under the choice of parameters $M<\frac{A^2-2D(E-C_L)}{2B}$ and $M^2<(E-C_L)^2$, the wave function is a sum of incoming and reflected contributions, 

\begin{eqnarray}
\Psi(x)&=&\Psi_i(x)+r\Psi_{r}(x)+\tilde{r}\Psi_{\tilde{r}}(x),
\end{eqnarray}
where $r$ ($\tilde{r}$) refers to the reflection amplitude of a propagating (evanescent) wave.

In the ($\left\vert E1\uparrow\right\rangle ,\left\vert
H1\uparrow\right\rangle$) basis and at given energy $E$ and transverse wave vector $k_y$, the two-component spinors can be written as

\begin{equation}
\Psi_i(x)=\left(
\begin{array}{c}
d(\mathbf{k})+M(\mathbf{k}) \\ 
A(k_{x}-ik_y)%
\end{array}%
\right) e^{ik_{x}x},
\label{psiincident}
\end{equation}

\begin{equation}
\Psi_{r}(x)=\left(
\begin{array}{c}
d(\mathbf{k})+M(\mathbf{k}) \\ 
-A(k_{x}+ik_y)%
\end{array}%
\right) e^{-ik_{x}x},
\label{psireflected}
\end{equation}

\begin{equation}
\Psi_{\tilde{r}}(x)=\left(
\begin{array}{c}
d(\mathbf{\kappa})+M(\mathbf{\kappa}) \\ 
-iA(\kappa_x+k_y)%
\end{array}%
\right) e^{\kappa_xx},
\label{prirevanescent}
\end{equation}
where $\mathbf{k}=(k_x,k_y)$, $k_x=\sqrt{k_1^2-k_y^2}$, and $\kappa_x=\sqrt{k_y^2+\kappa^2}$ with $\kappa_x>0$. 
In this Appendix, we use the definition $\kappa^2=-k_2^2$. The momenta $k_{1,2}$ are defined in Eq. (2). 

The wave function at the metal side reads
\begin{eqnarray}
\Psi(x)&=&t\Psi_{t}(x)+\tilde{t}\Psi_{\tilde{t}}(x),
\end{eqnarray}
with $t$ ($\tilde{t}$), the transmission amplitude of propagating (evanescent) modes and both transmitted wave functions are given by

\begin{equation}
\Psi_{t}(x)=\left(
\begin{array}{c}
d(\mathbf{k})+M(\mathbf{k}) \\ 
A(k_x-ik_y)%
\end{array}%
\right) e^{ik_xx},
\label{pritransmitted}
\end{equation}

\begin{equation}
\Psi_{\tilde{t}}(x)=\left(
\begin{array}{c}
d(\mathbf{\kappa})+M(\mathbf{\kappa}) \\ 
-iA(-\kappa_x+k_y)%
\end{array}%
\right) e^{-\kappa_xx}.
\label{psitevanescent}
\end{equation}
The momenta $k_x$ and $\kappa_x$ follow from Eq. (2) after the substitution $C_L \rightarrow C_R$. 
The spin down states are obtained from the previous ones Eqs. (2)-(4) and Eqs. (6) and (7) by substituting $k_x \rightarrow -k_x$.

The scattering amplitudes are obtained using scattering matrix theory by matching the left and right wave functions and their derivatives at the interface:

\begin{eqnarray}
\Psi(x,y)\vert_{x\rightarrow 0^-}&=&\Psi(x,y)\vert_{x\rightarrow 0^+},\nonumber\\
\partial_x\Psi(x,y)\vert_{x\rightarrow 0^-}&=&\partial_x\Psi(x,y)\vert_{x\rightarrow 0^+}.
\label{MatchingConditions}
\end{eqnarray}

In Sec. III. A, we investigate the behavior of the transmission probability and the amplitude squared of the evanescent modes for \textit{n-n'} and \textit{n-p} junctions.

The current conservation imposes the normalization of the transmission probability $|t|^2$ by the ratio of incoming and transmitted particle current, 
\begin{eqnarray}
|r|^2+|t|^2\Big|\frac{j_x^t}{j_x^i}\Big|&=&1,
\end{eqnarray}
where
\begin{eqnarray}
j_x^i&=&e\Psi_i^{\star}(x,y)\frac{\partial H}{\partial k_x}\Psi_i(x,y)\nonumber\\
&=&-2ek_1\cos\theta \{(D+B)(d(\mathbf{k})+M(\mathbf{k}))^2\nonumber\\
&+&A^2k_1^2(D-B)-A^2(d(\mathbf{k})+M(\mathbf{k}))\}.
\end{eqnarray}
The transmitted average current has the same form but depends on $C_R$ instead of $C_L$.
It is possible to solve the matching conditions and to find analytical expressions for all scattering amplitudes. However, the expressions are still too long to be written down here.

For fixed chemical potentials, Figs. $2$ and $3$ present the dependence of the transmission probability $T(\theta,M)=|t(\theta,M)|^2\big|\frac{j_x^t}{j_x^i}\big|$ and of the amplitude squared of evanescent modes $|\tilde{t}(\theta,M)|^2$ as functions of the incident angle $\theta$ and the mass parameter $M$.

\subsection*{2. HgTe/CdTe QW/superconductor interface}

We consider now the interface between a HgTe/CdTe QW and an $s$-wave superconductor. 
We assume the superconductivity is induced on the right-hand side of the interface by the proximity effect. 
Thus the effective Hamiltonian, Eq. (1), is expanded by particle-hole symmetry and contains the pairing potential $\Delta(x,y)=\Delta_0 e^{i\phi}$ ($\phi$ is the superconducting phase) as off-diagonal matrix elements, namely,

\begin{equation}
H_{BdG}=\begin{pmatrix}
h(-i \hbar \partial_{\mathbf{r}}) & \Delta(\mathbf{r}) \\
 \Delta^{\ast} (\mathbf{r}) &  -h(-i \hbar \partial_{\mathbf{r}})
\end{pmatrix},
\label{BDG}
\end{equation}
with $\mathbf{r}=(x,y)$.

The wave function on the left-hand side of the interface is the superposition of electron-like and hole-like quasiparticles,
\begin{eqnarray}
\Psi(x)&=&\Psi_i(x)+r_{ee}\Psi_{r_{ee}}(x)+\tilde{r}_{ee}\Psi_{\tilde{r}_{ee}}(x)\nonumber\\
&+&r_{he}\Psi_{r_{he}}(x)+\tilde{r}_{he}\Psi_{\tilde{r}_{he}}(x),
\end{eqnarray}
where $r_{ee}$ represents the amplitude of electrons to be reflected as electron-like quasiparticles at the interface while $r_{he}$ refers to the amplitude of an electron-hole conversion, namely an Andreev reflection process. The scattering amplitudes $\tilde{r}_{ee}$ and $\tilde{r}_{he}$ correspond to the reflection as electron- or hole-like evanescent modes.

By solving the Bogoliubov-de Gennes equation $H_{BdG}\Psi=E\Psi$, we obtain the expressions for the four-component vectors 
\begin{equation}
\Psi_{i}(x)=\left(
\begin{array}{c}
d(\mathbf{k})+M(\mathbf{k}) \\ 
A(k_x-ik_y)\\
0\\
0
\end{array}%
\right) e^{ik_xx},
\label{NSincident}
\end{equation}

\begin{equation}
\Psi_{r_{ee}}(x)=\left(
\begin{array}{c}
d(\mathbf{k})+M(\mathbf{k}) \\ 
-A(k_x+ik_y)\\
0\\
0
\end{array}%
\right) e^{-ik_xx},
\label{NSreflected}
\end{equation}

\begin{equation}
\Psi_{\tilde{r}_{ee}}(x)=\left(
\begin{array}{c}
d(\mathbf{\kappa})+M(\mathbf{\kappa}) \\ 
-iA(\kappa_x+k_y)\\
0\\
0
\end{array}%
\right) e^{\kappa_xx}.
\label{NSevareflected}
\end{equation}
The momenta $k_x=\sqrt{k_1^2-k_y^2}$ and $\kappa_x=\sqrt{\kappa^2+k_y^2}$ follow from Eq. (2). The propagating and evanescent hole-like wave functions 
follow from Eqs. (\ref{NSincident}) and (\ref{NSevareflected}) by the substitution $E\rightarrow -E$. Note that the electron-like excitations exist on 
the upper two components of the wave function and the hole-like excitations on the lower two components.

We focus on transport only below the superconducting gap ($E<\Delta_0$). Then the eigenfunctions on the superconducting side contain only evanescent waves

\begin{eqnarray}
\Psi(x)&=&t_{S+}\Psi_{S_+}(x)+t_{S-}\Psi_{S_-}(x)\nonumber\\
&+&t_{S'_+}\Psi_{S'_+}(x)+t_{S'_-}\Psi_{S'_-}(x),
\end{eqnarray}

where the spinors read

\begin{equation}
\Psi_{S_+}(x)=\left(
\begin{array}{c}
d(\mathbf{k}_S)+M(\mathbf{k}_S) \\ 
A(k_{xS}-ik_y)\\
\big[d(\mathbf{k}_S)+M(\mathbf{k}_S)\big]e^{i(\beta-\phi)}\\
A(k_{xS}-ik_y)e^{i(\beta-\phi)}
\end{array}%
\right) e^{(ik_{xS}-\xi)x},
\label{NSS+}
\end{equation}

\begin{equation}
\Psi_{S_-}(x)=\left(
\begin{array}{c}
d(\mathbf{k}_S)+M(\mathbf{k}_S) \\ 
A(k_{xS}-ik_y)\\
\big[d(\mathbf{k}_S)+M(\mathbf{k}_S)\big]e^{-i(\beta+\phi)}\\
A(k_{xS}-ik_y)e^{-i(\beta+\phi)}
\end{array}%
\right) e^{-(ik_{xS}+\xi)x},
\label{NSS-}
\end{equation}

\begin{equation}
\Psi_{S'_+}(x)=\left(
\begin{array}{c}
d(\mathbf{\kappa}_{S})+M(\mathbf{\kappa}_{S}) \\ 
-iA(-\kappa_{xS}+k_y)\\
\big[d(\mathbf{\kappa}_{S})+M(\mathbf{\kappa}_{S})\big]e^{i(\beta-\phi)}\\
-iA(-\kappa_{xS}+k_y)e^{i(\beta-\phi)}
\end{array}%
\right) e^{(\kappa_{xS}-i\xi)x},
\label{NSS'+}
\end{equation}

\begin{equation}
\Psi_{S'_-}(x)=\left(
\begin{array}{c}
d(\mathbf{\kappa}_{S})+M(\mathbf{\kappa}_{S}) \\ 
-iA(-\kappa_{xS}+k_y)\\
\big[d(\mathbf{\kappa}_{S})+M(\mathbf{\kappa}_{S})\big]e^{-i(\beta+\phi)}\\
-iA(-\kappa_{xS}+k_y)e^{-i(\beta+\phi)}
\end{array}%
\right) e^{(\kappa_{xS}+i\xi)x},
\label{NSS'-}
\end{equation}
and we have introduced the following parameters
\begin{eqnarray}
\beta&=&\arccos{\big(E/\Delta_0\big)},\\
k_{xS}&=&\sqrt{k_{S1}^2-k_y^2},\\
\kappa_{xS}&=&\sqrt{\kappa_{S}^2+k_y^2},\\
\xi&=&\frac{\sqrt{\Delta_0^2-E^2}}{\hbar v_F}.
\end{eqnarray}
The momenta squared $k^2_{S1}$ and $\kappa^2_S=-k^2_{S2}$ are defined by Eq. (2) after replacing $C_L$ by $C_R$ and setting $E$ to $0$. 

The scattering amplitudes follow from the matching conditions of wave functions and their derivatives at the interface as written in Eq. (\ref{MatchingConditions}). Their analytical expressions are long and therefore not presented here.

Also, due to current conservation, the Andreev reflection probability is normalized by the particle current of incoming electrons and reflected holes, namely,

\begin{eqnarray}
|r_{ee}|^2+|r_{he}|^2\Big|\frac{j_x^{hr}}{j_x^{ei}}\Big|&=&1,
\end{eqnarray}

where,
\begin{eqnarray}
j_x^{ei}&=&j_x(E)\nonumber\\
&=&-2ek_1\cos\theta \{(D+B)(d(\mathbf{k})+M(\mathbf{k}))^2\nonumber\\
&+&A^2k_1^2(D-B)-A^2(d(\mathbf{k})+M(\mathbf{k}))\}.
\end{eqnarray}
Due to particle-hole symmetry, the average current carried by the reflected hole is defined as $j_x^{hr}=j_x(-E)$.

Figure 4 shows the asymmetric behavior of the Andreev reflection probability $R_A(\theta,M)=|r_{he}(\theta,M)|^2\big|\frac{j_x(-E)}{j_x(E)}\big|$ as a function of the angle of incidence and the mass term and its correlation with the non-zero amplitude squared of hole-like evanescent modes $|\tilde{r}_{he}(\theta,M)|$.



\begin{thebibliography}{99}

\bibitem{Dyakonov1971} M. I. Dyakonov and V.I. Perel, Phys. Lett. A \textbf{35}, 459 (1971).
\bibitem{Hirsch1999} J. E. Hirsch, Phys. Rev. Lett. \textbf{83}, 1834 (1999).
\bibitem{Murakami2003} S. Murakami, N. Nagaosa, and S.-C. Zhang, Science \textbf{301}, 1348 (2003).
\bibitem{Sinova2004} J. Sinova, D. Culcer, Q. Niu, N. A. Sinitsyn, T. Jungwirth, and A. H. MacDonald, Phys. Rev. Lett. \textbf{92}, 126603 (2004).
\bibitem{Kato2004} Y. Kato, R. C. Myers, A. C. Gossard, and D. D. Awschalom, Science \textbf{306}, 1910 (2004).
\bibitem{Garlid2010} E. S. Garlid, Q. O. Hu, M. K. Chan, C. J. Palmstrom, and P. A. Crowell, arXiv:1006.1163.
\bibitem{Wunderlich2005} J. Wunderlich, B. Kaestner, J. Sinova, and T. Jungwirth, Phys. Rev. Lett. \textbf{94}, 47204 (2005).
\bibitem{Brune2010} C. Br\"{u}ne, A. Roth, E. G. Novik, M. König, H. Buhmann, E. M. Hankiewicz, W. Hanke, J. Sinova, and L. W. Molenkamp, Nature Phys. \textbf{6}, 448 (2010).
\bibitem{Brune2011} C. Br\"{u}ne, A. Roth, H. Buhmann, E. M. Hankiewicz, L. W. Molenkamp, J. Maciejko, X.-L. Qi, and S.-C. Zhang, arXiv:1107.0585.
\bibitem{Konig2007} M. K\"{o}nig, S. Wiedmann, C. Br\"{u}ne, A. Roth, H.
Buhmann, L. Molenkamp, X.-L. Qi, and S.C. Zhang, Science {\bf 318}, 766 (2007).
\bibitem{Kane2005} C. L. Kane and E. J. Mele, Phys. Rev. Lett. \textbf{95}, 226801 (2005).
\bibitem{Bernevig2006} B. A. Bernevig, T. L. Hughes, and S.-C. Zhang, Science \textbf{314}, 1757 (2006).
\bibitem{Fu2007} L. Fu and C. L. Kane, Phys. Rev. B \textbf{76}, 045302 (2007).\
\bibitem{Buttner2010} B. B\"{u}ttner, C.-X. Liu, G. Tkachov, E. G. Novik, C. Br\"{u}ne, H. Buhmann, E. M. Hankiewicz, P. Recher, B. Trauzettel, S.-C. Zhang, and L.W. Molenkamp, Nature Phys. {\bf 7}, 418 (2011).
\bibitem{Yokoyama2009} T. Yokoyama, Y. Tanaka, and N. Nagaosa, Phys. Rev. Lett. \textbf{102}, 166801 (2009).
\bibitem{Novik2005} E. G. Novik, A. Pfeuffer-Jeschke, T. Jungwirth, V. Latussek, C. R. Becker, G. Landwehr, H. Buhmann, and L. W. Molenkamp, Phys. Rev. B {\bf 72}, 035321 (2005).
\bibitem{Tworzydlo2006} J. Tworzydlo, B. Trauzettel, M. Titov, A. Rycerz, and C. W. J. Beenakker, Phys. Rev. Lett. {\bf 96}, 246802 (2006).
\bibitem{Beenakker2006} C. W. J. Beenakker, Phys. Rev. Lett. {\bf 97}, 067007 (2006).
\bibitem{Novik2009} E. G. Novik, P. Recher, E. M. Hankiewicz, and B. Trauzettel, Phys. Rev. B \textbf{81}, R241303 (2010).
\bibitem{LBZhang2009} L. B. Zhang, K. Chang, X. C. Xie, H. Buhmann, and L. W. Molenkamp, New J. Phys. \textbf{12}, 083058 (2010).
\bibitem{foot_rothe} The origin of this effect might be indirectly related to the in-plane Pauli term discovered by D. G. Rothe, R. W. Reinthaler, C.-X. Liu, L. W. Molenkamp, S.-C. Zhang, and E. M. Hankiewicz, New. J. Phys. {\bf 12}, 065012 (2010).
However, these authors work in a different validity regime of physical parameters than do.
\bibitem{Guigou2010} M. Guigou and J. Cayssol, Phys. Rev. B \textbf{82}, 115312 (2010).
\bibitem{Yamakage2010} A. Yamakage, K.-I. Imura, J. Cayssol, and Y. Kuramoto, Phys. Rev. B \textbf{83}, 125401 (2011).
\bibitem{Gao2011}J.-H. Gao, J. Yuan, W.-Q. Chen, Y. Zhou, and F.-C. Zhang, Phys. Rev. Lett. {\bf 106}, 057205 (2011).
\bibitem{Werake2011} L. K. Werake, B. A. Ruzicka, and H. Zhao, Phys. Rev. Lett. {\bf 106}, 107205 (2011).


\end{thebibliography}
\end{document}